\documentclass[10pt,twocolumn,journal]{IEEEtran}  

\IEEEoverridecommandlockouts                              

\usepackage{srcltx} 
\usepackage{amsmath,amsxtra,amssymb,amsthm}
\usepackage{times,latexsym,array,verbatim,url}
\usepackage{color,graphicx,cite}
\usepackage{datetime}

\newtheorem{tips}{Suggestion}
\newtheorem{temp}{Template}
\newtheorem*{tips*}{Suggestion}
\newtheorem*{temp*}{Template}

\theoremstyle{remark}

\title{\LARGE \bf A Template and Suggestions for Writing \\ Easy-to-Read Research Articles}

\author{Tansu Alpcan
\thanks{Department of Electrical and Electronic Engineering,
 The University of Melbourne, Australia. Email: $tansu.alpcan@unimelb.edu.au$.} 
}

\begin{document}

\maketitle
\thispagestyle{empty}
\pagestyle{empty}

\begin{abstract}
The number of research papers written has been growing at least linearly -- if not exponentially --
in recent years. In proportion, the amount of time a reader allocates per paper has been decreasing. 
While an accessible paper will be appreciated by a large audience, hard-to-read papers may remain obscure
for a long time regardless of scientific merit. Unfortunately, there is still insufficient
emphasis on good written and oral communication skills in technical disciplines, especially in engineering.

As an academic, I have realised over the years that I keep telling my students the same things over and over again when they write papers, reports, presentations, and theses. This article contains some of those suggestions and serves as a limited template for organising research articles. I have adopted a very practical and personal approach and don't claim that this is a formal contribution to the scientific communication literature. However, I hope that this article will not only make my life a bit easier but also help other graduate students and academic supervisors.
\end{abstract}


\section*{Preliminaries} \label{sec:notation}

This paper serves as a template, contains suggestions, and discusses properties of well-written and well-organised papers.
The templates for the specific sections are formatted as:
\begin{temp*}[Section] 
Here is one way to organise this section:
 \begin{enumerate}
  \item It should contain such and such.
 \end{enumerate}
\end{temp*}
Specific suggestions are emphasised using the following:
\begin{tips*}[Section]
A few relevant tips and tricks:
 \begin{itemize}
  \item use this approach when writing this section.
 \end{itemize}
\end{tips*}

The plain text parts of the paper discuss the best practices, present resources, and contain the author's opinions.

\section{Introduction} \label{sec:intro}

I think almost all of my papers begin with a section titled ``Introduction''. It makes a lot of sense to start the paper by introducing the reader what the paper is about. 

Many authors fail to realise that the readers do not really know anything about the author's research. Specifically, students often assume that the reader has the necessary background knowledge (including acronyms), appreciates the underlying problem they work on, and will fill in the blanks when presented the results. The reality is of course the opposite. Most of the time, the reader barely knows the topic and the methods, has no idea why the presented work is significant or useful, and cannot make sense of the results unless the implications are explicitly discussed and clarified.

It should be also crystal clear to any author that the first page of a paper is a ``prime resource'' in our 
time-constrained `information age'. The first page is the face of the paper and as we all know, the first impressions matter a lot in human psychology. Unfortunately, many authors waste the half of the first page with banal generalities and filler sentences instead of using it efficiently.

\begin{temp}[Introduction] 
A good way to organise the Introduction Section is as follows:
\begin{enumerate}
   \item The author should first explain briefly the domain and the main theme(s) of the paper. What is this paper about? Most journals and conferences ask for a few keywords, which may provide a good clue.
  \item Next, the author should describe the motivation, introduce the main research question(s), and argue for their significance. What is the question this paper tries to answer and why should the reader care?
  \item The contributions presented in the paper and their novelty should be explained explicitly. What is done here that was not done before in the literature? Is it the model, theoretical contributions, simulations, experiments, or a combination of these? A brief comparison to prominent existing works is absolutely necessary.
  \item Space permitting, it is a good idea to have a subsection called ``Contributions'' and explicitly list the main novelty and contributions of the paper for improved readability.
  \item It is often customary to give an overview of the paper organisation in a single paragraph, explicitly listing the remaining sections and what they are about.
\end{enumerate}
\end{temp}

\begin{tips}[Introduction]
A few relevant tips and tricks for the Introduction Section.
 \begin{itemize}
  \item Ideally, the first three points in the template should fit to the first page. Hence, the reader gets an overview of the topic, the question(s) addressed, why these research questions are important, and what are the main contributions of the paper in comparison to the literature, all from the first page.
  \item The introduction should focus on what the authors have done and how (methodology). A common mistake is spending too many words in the Introduction on irrelevant generalities, background that does not further the main story, or the contributions of other works.
  \item The introduction sets the tone for the upcoming discussions later in the result sections. If it is boring and confusing, most technical readers jump to the model section quickly without having a good idea about what problem the paper aims to address and why.
 \end{itemize}
\end{tips}

\section{Literature Review} \label{sec:litreview}

No scientific work exists in vacuum. Even the most innovative ideas have a connection to the existing literature. In fact, most of the scientific papers as of early 21st century are rather incremental in nature. The main job of the literature review section is to position the paper's contributions within the literature. It also indicates the reader that the authors have done their homework and checked what is already out there before claiming novelty of their contributions. If there is space, the literature review deserves its own section. If not, e.g. in conference papers, it can be embedded into the introduction section. 

A common mistake the students do in this section is to summarise each relevant paper they have read as part of the project in couple of sentences without any organisation or giving the reader any insights. Ideally, this section should be organised just like a mini literature survey paper.

\begin{temp}[Literature Review] 
Using the list of the contributions from the Introduction, start with a paragraph mentioning the relevant background to each contribution item. This leads to a list of relevant topics that were previously explored by others. Instead of a dry summary, the goal is to familiarise the reader with what was done before and \underline{highlight the gaps in the literature}. It may look like this:
\begin{enumerate}
  \item Starting paragraph: the topic has multiple interconnected aspects, A1-A3, which were explored before in the literature...
  \item A1 was discussed early on in [1]. Another work [2] presented a novel \textit{xxx}. However, \textit{yyy} was not explored before.
  \item Another important aspect is A2. The rich literature on A2 is summarised in the survey paper [3]. 
  \item A3 was proposed by [4] and extended further by [5]. 
  \item However, no paper combined A1 and A2 and extended it to this new direction \textit{zzz} to the best of our knowledge.
\end{enumerate}
\end{temp}

\begin{tips}[Literature Review]
Good organisation is the key to a good and readable literature review section.
 \begin{itemize}
  \item Avoid listing one paper after another without any conceptual organisation.
  \item Only mention the works that are directly relevant to the paper themes. It is good to cite authoritative books and survey papers to save space and help the reader.
  \item A good literature review should highlight the gaps, clarify the contributions, and help the reader understand the position of the paper in the grand scheme of things.
 \end{itemize}
\end{tips}

\section{Problem Formulation} \label{sec:problem}

If there is sufficient space, it is good to dedicate a section to formulate the problems discussed in the paper.
The title of this section does not have to be ``Problem Formulation.'' Actually, it is often better to use a descriptive title from the specific problem domain. This is the part where individual papers start to diverge from each other. If the problem is well-known, then this section can be embedded into the model section or the preceding introduction section. If it is a new or complex problem, then it may need to be explained to the reader in detail. The problem should also be properly motivated. This section should make the significance of the problem crystal clear to the reader. 

\begin{temp}[Problem Formulation] 
The overarching problem this paper address is \textit{xxx}, which can be decomposed into:
\begin{enumerate}
  \item How to achieve yyy as a preliminary step?
  \item What is the best method for \textit{zzz}, so that together with \textit{yyy}, \textit{xxx} is addressed satisfactorily, i.e. the metric \textit{W} is maximised under constraints \textit{Y}.
  \item Solving problem \textit{xxx} has the following practical implications: ...
\end{enumerate}
\end{temp}

\begin{tips}[Problem Formulation] 
This section is either embedded into one the others (model or introduction) or is closely connected to them.
 \begin{itemize}
  \item A brief summary of the problem should be on the first page of the paper (introduction) to prepare the reader to this extended version.
  \item If the problem arises within a specific model, then the model section should come first or the problem can be embedded into the end of the model section as a subsection.
  \item Many technical papers do not bother to explain the reader why this problem is important and what solving it will achieve. Pure mathematicians may not care how their art is used in practice but everyone else, e.g. engineers should tell the readers explicitly what solving this problem will bring.
 \end{itemize}
\end{tips}

\section{The Model} \label{sec:model}

Modelling is at the heart of modern science. Therefore, it is not surprising that most papers rely on some type of a model. The model could be mathematical or algorithmic or simulation-based or experimental, or a combination of these. The section title can be chosen accordingly, e.g. it could named simulation or experimental setup instead of the model. Again, descriptive titles are much more preferable over generic ones.

The model used in the paper is the foundation of the paper's contributions. If the model is not explained well and the readers are left on a shaky foundation, they would naturally neither understand, nor appreciate the hard work of the authors.

\begin{temp}[Model] 
Most papers build upon a lot of background knowledge and it is impossible to include all of it in a paper. Still, the author should try to provide the reader the material and pointers for completeness as much as the limited space permits.
\begin{enumerate}
  \item It is good to point out and give preliminary knowledge with couple of sentences and good references. This tells the knowledgeable readers where the authors come from and less knowledgeable ones clues about what background is needed to understand the model. 
  \item The model should be explained briefly and clearly. It should be supported by relevant references. 
    \begin{itemize}
     \item Mathematical: the relevant equations upon which the later sections build.
     \item Algorithmic: describing the broad class of algorithms, which the contributed ones belong.
     \item Simulation-based: the simulator, maybe why it was chosen with a sentence, the input data, high-level simulation setup.
     \item Experimental: the experiment setup, the hardware and software used, data...
    \end{itemize}
   \item It is very important to highlight and discuss the underlying assumptions of the model (or limitations of the simulation/data/experiments). Hiding important assumptions made in the model is dishonest and bad practice. 
   \item If desired, a subsection on the approach that is used in the paper could be added to the end of the model section. They are different things, so a subsection header is needed to differentiate. Alternatively, the approach can be embedded to the (beginning of) result sections.
\end{enumerate}
\end{temp}

\begin{tips}[Model] 
The model section often does not present the paper's contributions but lays the foundation on which they stand. 
 \begin{itemize}
  \item Many experienced readers jump to the model section very quickly after a quick glance to the first page. For expert readers, the model section is the face of the paper. 
  \item Given that most papers are published in specialised journals and conferences, it is natural to assume that readers have some background on the topics of the paper. The model section should be tailored to the readership. Something that is obvious to one set of readers could be mysterious to another depending on the research community. For example, the same contribution may need a totally different model (and background) section for different venues. This is especially tricky in interdisciplinary research. 
  \item The model is different from the approach used in the paper. The approach subsection should smoothly bridge the model and the result sections, regardless of being placed with the former or the latter.
  \item Some researchers (unfortunately not too few) think that by not clarifying the assumptions and limitations of their models, they can oversell their results. I find this very unethical and not so clever. Remember, as an author, one can fool many people some of the time and some people all the time, but not all the people all the time. Furthermore, all papers are archived for perpetuity. How will it look like ten or twenty years later?
 \end{itemize}
\end{tips}

\section{The Result Sections} \label{sec:results}

Before presenting the \textit{beautiful and ground-breaking} results of the paper, it may be a good idea to briefly explain the approach used to obtain them. Why not include an approach subsection or a paragraph or two, which clearly explain what the authors did to obtain the results and how?

The results sections are the heart of the paper and contain the main contributions. Again, these contributions may be mathematical, algorithmic, simulation-based, experimental, or a mixture of these. The results should be presented over multiple sections in a well-organised paper. 

It is hard to be prescriptive about the results sections due to the diversity of research contributions. However, I would like to make some suggestions based on past experience.

\begin{tips}[Results] 
Whether solving the important problem described previously or introducing a novel methodology, the results sections present the main content. In fact, everything else in the paper is merely support material. The presentation should be organised carefully to communicate the research contributions to the reader in an easy-to-follow structure.
\begin{itemize}
  \item Once a set of results are obtained, it is a good idea to take a step back and decide which results will be presented in the paper. This requires a judgement on significance, which is not easy. 
    \begin{itemize}
     \item If there are too few results, then clearly additional work is needed; maybe further analysis of different aspects of the problem or solution. 
     \item If there is too much material, hard choices need be made on what to include within the allocated page limit, i.e which results are really important. These days, one can always upload a longer version of the work to a repository such as Arxiv to provide more details.
    \end{itemize}
  \item Once the \textit{set} of results is determined, that set needs to be converted to an \textit{ordered set}. The paper is necessarily in a linear format so the author has to play the role of story teller to the reader, explaining the results one by one in an organised and logical way. This is easy if the author is clear on motivation and problem formulation. Most students (and some authors) struggle telling a story because they are confused about those two, i.e. why they do what they do. In that case, the supervisors (or senior co-authors) can give the much needed support. 
  \item Ideally, the paper should have a discussion section, subsection or paragraphs. Once the results are properly conveyed to the reader, they need to be interpreted and discussed. Many technical papers fail to accomplish this task. What do these results mean and what are their implications? These two points need to be explicitly explained and crystal clear. If the authors cannot achieve this maybe the authors themselves are confused and how can they expect the reader not to be? 
  \item Visualisation is very important. We live in a visual century and a picture has always been worth a thousand words. Specific tips:
   \begin{itemize}
    \item Graphs and bar graphs are better than tables, which often belong more to appendices. If a table has to be used, the important cells and columns/rows should be shaded/highlighted.
    \item A graph should have a good title, legible axes labels, and a descriptive caption that stands on its own, e.g. ``\textit{Algorithms A, B, C are compared under Y conditions for various Q. Curve A describing the output metric X of Algorithm A clearly outperforms Curve B of Algorithm B as quantity Q increases. etc. etc.}''
    \item Each graph should tell a story and highlight an important result. Just because a graphical result exists does not mean it is worth including in the paper (and in that form). Many results can be told verbally within the text if the contributions are trivial or as expected. Again, a link to the extended version can be provided to the reader.
   \end{itemize}
\end{itemize}
\end{tips}

\section{Conclusions} \label{sec:conclusion}

This section is very similar to the abstract and introduction but also has a lot of advantages over them. Now that the reader has seen all the results, one can provide a much more informed summary that contains insights. The conclusion section should convey the main points of the entire paper and the bottom line of the work. This obviously requires the authors achieving great clarity in their understanding of their own work.

\begin{temp}[Conclusions] 
This section may be thought like a mini paper condensing the entire work, followed by the future directions.
\begin{enumerate}
  \item Background, motivation, problem addressed.
  \item The model, approach, and methods.
  \item A verbal summary of the main results, discussion, and implications/importance.
  \item Future directions, e.g. what did not fit to this paper, could have done if time/resources permitted, or what results would have been nice to have based on the insights of this paper.
\end{enumerate}
\end{temp}


\section{Abstract} \label{sec:abstract}

It may look paradoxical but makes a lot of sense to write the abstract \textbf{chronologically} at the very end. By that time, the authors hopefully have a very clear idea what the paper is about and its messages. Here is a suggested template for the abstract:

\begin{temp}[Abstract] 
Many abstracts in engineering articles are deadly boring. In contrary, the abstract should excite and motivate the readers to read the rest of the paper!
\begin{enumerate}
  \item One or two sentences on background and motivation.
  \item A sentence on the problem addressed.
  \item Couple of sentences on the model, approach, and methods; their novelty in contrast to existing works.
  \item Highlights of exciting main results in two-three sentences. 
  \item A closing sentence on the implications/importance of the contributions/results.
\end{enumerate}
\end{temp}

\section*{Tips and Additional Resources} \label{sec:tips}

\subsection*{Technical Writing}

English is today the de-facto language of science. Using this language properly is challenging for authors whose native language is not English (like myself). Clear writing requires, however, more than the expertise on the rules and mechanics of a specific language. Achieving the clarity appropriate for a good scientific publication requires honesty, clear thinking, and maturity. The following three books may help in this endeavour:

\textbf{1.} J. Zobel, ``Writing for Computer Science''. Springer London, 2015.

This book~\cite{zobel2015writing} has Computer Science in its title but should help engineers as well. It covers the fundamentals very well. Highly recommended.

\textbf{2.} W. Strunk, E. White, and M. Kalman, ``The Elements of Style''. A Penguin book : Reference, Penguin Books, 2007.

This~\cite{strunk2007elements} is an absolute classic on the nuts and bolts of writing. A must read.

\textbf{3.} J. Williams and G. Colomb, ``Style: Toward Clarity and Grace''. Chicago guides to writing, editing, and publishing, University of Chicago Press, 1995.

This~\cite{williams1995style} is the book to read once the author has a good understanding of the basics. It is one of my favourites.

\subsection*{References}

No scientific work exists in vacuum. The references should be used appropriately  
\begin{itemize}
 \item as pointers to the relevant background work.
 \item as part of the literature review. The cited papers create the landscape in which the paper lives. The editors often use the authors of the related cited papers as a peer-review resource. This practical aspect should also be taken into account!
 \item to point directions and works that go beyond the scope of the paper.
\end{itemize}

The references should be appropriately formatted. Not following the formatting rules, e.g. of IEEE or others as required by the publication venue, looks very sloppy and gives the impression that the authors are not professionals.
It is easy to find citations online but many repositories' formats differ from the required one, so it still requires some post-processing.

\subsection*{Final Comments}

Additional general tips and comments:
\begin{itemize}
 \item Always prefer explicit over implicit. Don't expect the reader to come to conclusions and fill in the blanks; provide them clear material and your conclusions. Don't worry, the readers will make up their own mind anyway once they understand the material. The problem is when they have a foggy understanding of it.
 \item Always use a spell checker!
 \item Latex looks nicer than word processors but requires a significant upfront investment. 
 \item Block diagrams, algorithms, and graphics that illustrate the concepts are very very good and should be used liberally throughout the paper. One picture is worth thousand words.
 \item Scalable vector graphics (svg) or pdf, eps, should be preferred over bitmap (png or jpg). Consider trying \textit{Inkscape}, which is a good open source multi-platform program.
 \item A \textit{scannable pape}r is much more accessible than one that is hard to understand. It roughly means: the reader should get a very good idea about the paper's topic and contributions just by looking at the first page, the graphs/graphics with their captions, section titles, and conclusion.
\end{itemize}

\section*{Presentation Checklist}

Many scientists are quite bad in making presentations. I cannot count the number of boring, incomprehensible presentations I have suffered personally in scientific conferences. Things are improving slowly with newer generations but still...

This checklist is based on a mini course I took from Deutsche Telekom in Germany years ago. Unfortunately, I forgot the name of the instructor so cannot give him the credit he deserves for the nice job he had done.

\textbf{Specify the cause of the presentation}
\begin{itemize}
 \item Who proposed the presentation? \underline{Why}?
 \item Which \underline{significance} does the presentation have from the point of view of xxx (in audience)?
\end{itemize}

\textbf{Define the objectives and goals of your presentation}
\begin{itemize}
 \item What are the factual and emotional \underline{objectives} that I want to achieve?
\end{itemize}

\textbf{Analyse the expected audience}
\begin{itemize}
 \item Which \underline{expectations}, needs, wishes does my audience have?
 \item How will the \underline{audience benefit} from my presentation?
 \item What are the main properties of the audience (\underline{background knowledge}, education)? 
 \item Which prejudices may prevail?
\end{itemize}

\textbf{Define the content}
\begin{itemize}
 \item What are the \underline{key messages} (Max. 3)?
 \item What is the extend of \underline{background information} needed?
 \item Which \underline{examples} enforce the effect?
 \item What can be omitted?
\end{itemize}

\textbf{Structure your presentation}
\begin{itemize}
 \item In the introduction (“catch them”; 15\%) 
 \item The main part (“keep them”; 75\%)
 \item Within the final part („convince them“; 10\%) 
 \item Pyramid Structure: does each slide have a message?
\end{itemize}

\textbf{Define the \textit{storybook}}
\begin{itemize}
 \item \underline{Jokes}, exaggerations, getting to the point: using pictures, quotations ...
 \item Make I use of my \underline{own stories} or ones from the experiences of my audience?
 \item Do I use lively, \underline{active speech}?
 \item Do I present a change in \underline{presentation style} every 10 minutes?
\end{itemize}

\textbf{Work on presentation style}
\begin{itemize}
 \item How do I use my \underline{voice and body language} ?
 \item Are there enough \underline{short breaks} in my presentation? (The art of making a pause)
 \item Am I \underline{personally convinced} and glad to give my audience something valuable?
 \item Do I show my personal affection to the topic?
 \item Do I have eye contact with the audience?
\end{itemize}

He who presents is in the focus of the auditorium!

Slides may be helpful but only as an auxiliary means!

\section*{Acknowledgement}                               
The author wishes to thank his current and past PhD students and postdocs for their feedback, comments, and suggestions.


\begin{thebibliography}{1}
\providecommand{\url}[1]{#1}
\csname url@samestyle\endcsname
\providecommand{\newblock}{\relax}
\providecommand{\bibinfo}[2]{#2}
\providecommand{\BIBentrySTDinterwordspacing}{\spaceskip=0pt\relax}
\providecommand{\BIBentryALTinterwordstretchfactor}{4}
\providecommand{\BIBentryALTinterwordspacing}{\spaceskip=\fontdimen2\font plus
\BIBentryALTinterwordstretchfactor\fontdimen3\font minus
  \fontdimen4\font\relax}
\providecommand{\BIBforeignlanguage}[2]{{%
\expandafter\ifx\csname l@#1\endcsname\relax
\typeout{** WARNING: IEEEtran.bst: No hyphenation pattern has been}%
\typeout{** loaded for the language `#1'. Using the pattern for}%
\typeout{** the default language instead.}%
\else
\language=\csname l@#1\endcsname
\fi
#2}}
\providecommand{\BIBdecl}{\relax}
\BIBdecl

\bibitem{zobel2015writing}
\BIBentryALTinterwordspacing
J.~Zobel, \emph{{Writing for Computer Science}}.\hskip 1em plus 0.5em minus
  0.4em\relax Springer London, 2015. [Online]. Available:
  \url{https://books.google.com.au/books?id=LWCYBgAAQBAJ}
\BIBentrySTDinterwordspacing

\bibitem{strunk2007elements}
\BIBentryALTinterwordspacing
W.~Strunk, E.~White, and M.~Kalman, \emph{{The Elements of Style}}, ser. {A
  Penguin book : Reference}.\hskip 1em plus 0.5em minus 0.4em\relax Penguin
  Books, 2007. [Online]. Available:
  \url{https://books.google.com.au/books?id=sj5\_wr6zIEcC}
\BIBentrySTDinterwordspacing

\bibitem{williams1995style}
\BIBentryALTinterwordspacing
J.~Williams and G.~Colomb, \emph{{Style: Toward Clarity and Grace}}, ser.
  {Chicago guides to writing, editing, and publishing}.\hskip 1em plus 0.5em
  minus 0.4em\relax University of Chicago Press, 1995. [Online]. Available:
  \url{https://books.google.com.au/books?id=wheCdBUIpwIC}
\BIBentrySTDinterwordspacing

\end{thebibliography}


\end{document}